\documentclass[times,twocolumn,final]{elsarticle}
\usepackage{framed,multirow}
\usepackage{amsmath}
\usepackage{amssymb}
\usepackage{latexsym}
\usepackage{url}
\usepackage{xcolor}
\usepackage{hyperref}
\usepackage{booktabs}
\usepackage{color}
\usepackage{geometry}
\graphicspath{{./Figs/}}

\def\x{{\bf x}}

\def\0{{\bf 0}}
\def\1{{\bf 1}}

\def\etal{{\em et al.}}
\def\eg{{\em e.g.}}
\def\ie{{\em i.e.}}

\def\etal{{\em et al.\/}\,}

\definecolor{newcolor}{rgb}{.8,.349,.1}

\begin{document}
\begin{frontmatter} 
\title{Leveraging Brain Modularity Prior for Interpretable Representation Learning of fMRI\\
} 
\author{Qianqian Wang}
\author{Wei Wang}
\author{Yuqi Fang}
\author{P.-T. Yap}
\author{Hongtu Zhu}
\author{Hong-Jun Li\corref{cor1}}
\author{Lishan Qiao}
\author{Mingxia Liu\corref{cor1}}

\cortext[cor1]{Corresponding authors:   M.~Liu (mxliu@med.unc.edu) and H.~Li (lihongjun00113@ccmu.edu.cn).}

\begin{abstract}
Resting-state functional magnetic resonance
imaging (rs-fMRI) can reflect spontaneous neural activities in brain and is widely used for brain disorder analysis.
Previous studies propose to extract fMRI representations through diverse machine/deep learning methods for subsequent analysis.   
But the learned features typically lack biological interpretability, 
which limits their clinical utility. 
\if false
Previous studies have designed diverse machine/deep learning methods to extract latent fMRI representation, 
but the learned representation typically lacks biological interpretability, 
which limits their clinical practicability.
\fi 
%
From the view of graph theory,
the brain exhibits a remarkable modular structure
in spontaneous brain functional networks, with each module comprised of functionally interconnected brain regions-of-interest (ROIs).
However, most existing learning-based methods for fMRI analysis fail to 
adequately utilize such brain modularity prior.
In this paper,
we propose a \textbf{B}rain \textbf{M}odularity-constrained dynamic \textbf{R}epresentation learning (BMR) framework for interpretable fMRI analysis, 
consisting of three major components:
(1) dynamic graph construction, 
(2) dynamic graph learning via a novel modularity-constrained graph neural network (MGNN), 
(3) prediction and biomarker detection for interpretable fMRI analysis. 
Especially, three core neurocognitive modules (\ie, salience network, central executive network, and default mode network) are explicitly incorporated into the MGNN, encouraging the nodes/ROIs within the same module to share similar representations. 
To further enhance discriminative ability of learned features, we also encourage the MGNN to preserve the network topology of input graphs via a graph topology reconstruction constraint. 
Experimental results on 534 subjects with rs-fMRI scans from two datasets validate the effectiveness of the proposed method. 
The identified discriminative brain ROIs and functional connectivities can be regarded as potential fMRI biomarkers to aid in clinical diagnosis. 
\end{abstract}
\begin{keyword}
Functional MRI \sep Brain modularity \sep Brain disorder \sep Biomarker.
\end{keyword}
\end{frontmatter}


\begin{figure*}[t]
\setlength{\abovecaptionskip}{0pt}
\setlength{\belowcaptionskip}{0pt}
\setlength{\abovedisplayskip}{0pt}
\setlength{\belowdisplayskip}{0pt}
\centering
\includegraphics[width=1\textwidth]{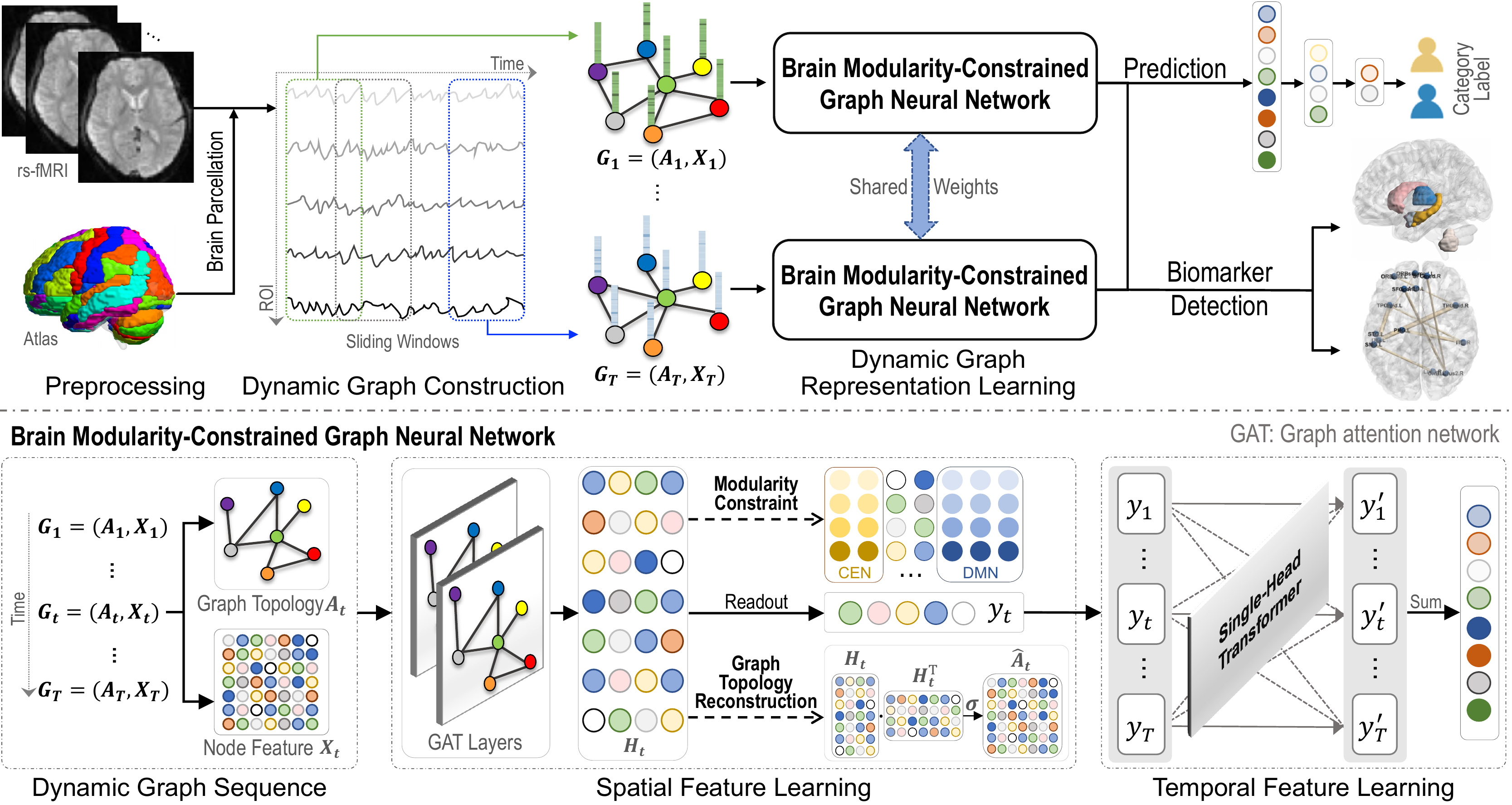}
\caption{Illustration of the proposed Brain Modularity-constrained dynamic
Representation learning ({BMR}) framework for interpretable
functional MRI analysis, including three major components: 
(1) dynamic graph construction using sliding windows, 
(2) dynamic graph representation learning via a novel modularity-constrained graph neural network (MGNN), and 
(3) prediction and biomarker detection for interpretable brain disorder analysis. 
The BMR is constrained by three fundamental neurocognitive modules, \ie, salience network (SN), central executive network (CEN), and default mode network (DN). 
To further enhance discriminative ability of learned fMRI representations, we also encourage BMR to reconstruct network topology of input graphs.}
\label{pipeline}
\end{figure*}
\section{Introduction}
\label{S1}
Resting-state functional magnetic resonance imaging (rs-fMRI) provides a noninvasive solution to reveal brain spontaneous neural activities by measuring blood-oxygenation-level-dependent (BOLD) signals~\cite{bolt2022parsimonious,pervaiz2022multi,sip2023characterization}.
It has been increasingly used to understand underlying neuropathological mechanisms of various brain disorders, such as autism spectrum disorder and cognitive impairment~\cite{li2021braingnn,kunda2022improving,dsouza2019multivoxel}.
Many machine/deep learning-based methods have been proposed to map 4D fMRI data into low-dimensional representations
and perform downstream brain disease detection~\cite{nebel2022accounting,azevedo2022deep,bessadok2022graph}.
However, due to the complexity of brain organization and the black-box property of many learning-based methods,
the generated fMRI representations usually lack 
biological interpretability, 
thereby limiting their clinical practicability~\cite{geirhos2020shortcut,chen2022explainable}. 

From the perspective of graph theory, the human brain 
exhibits a significant modular structure in spontaneous brain functional networks (BFN), with each module executing specialized cognitive function~\cite{he2009uncovering}.
A functional module can be defined as a subnetwork of densely interconnected brain regions-of-interest (ROIs) that are sparsely connected to ROIs in other modules~\cite{sporns2016modular,meunier2010modular}.
In particular, 
salience network (SN), central executive network (CEN), and default mode network (DMN) are three fundamental neurocognitive modules/subnetworks in human brains, supporting effective cognitive activities~\cite{goulden2014salience}.
Unfortunately, existing fMRI-based studies usually 
fail to adequately utilize such brain modularity prior 
during fMRI representation learning.
On the other hand, 
the brain can be modeled as a spatiotemporally dynamic BFN (\ie, dynamic graph) based on BOLD signals, aiming to 
help simultaneously capture spatial and temporal information of brain neural activities~\cite{kong2021spatio,gadgil2020spatio}.
Intuitively, it is meaningful to incorporate brain modularity prior into a dynamic graph representation framework for interpretable fMRI analysis.


To this end, we propose a Brain Modularity-constrained dynamic
Representation learning (\textbf{BMR}) framework for interpretable
fMRI analysis.
As shown in the top panel of Fig.~\ref{pipeline}, the proposed BMR consists of three major components: (1) dynamic graph construction using sliding windows, 
(2) dynamic graph representation learning via a novel brain modularity-constrained graph neural network (MGNN), 
and (3) prediction and biomarker detection for interpretable fMRI analysis. 
In particular, three core neurocognitive modules (\ie, SN, CEN, and DMN) are explicitly incorporated into the BMR through our proposed modularity constraint, encouraging learned features of ROIs within the same module to be similar. 
To enhance discriminative ability of learned features, we design a graph topology reconstruction constraint to encourage our BMR to preserve the network topology of input graphs during fMRI feature learning.  
Experimental results on 534 subjects with rs-fMRI from the public Autism Brain Imaging Data Exchange (ABIDE) dataset~\cite{di2014autism} and a private HIV-associated neurocognitive disorder (HAND) dataset 
demonstrate the superiority of the proposed BMR over several state-of-the-art methods for brain disorder detection.

The contributions of this work are summarized as follows.
\begin{itemize}
\item
A novel brain modularity-constrained dynamic
representation learning framework is designed for interpretable
fMRI analysis, where brain ROIs within the same functional module are encouraged to share similar representations. 
To the best of our knowledge, this is among the first attempts to incorporate brain modularity prior into graph neural networks for fMRI analysis.  
\item
A graph reconstruction constraint is introduced during dynamic graph learning to preserve original topology information of brain functional networks,  thus enhancing discriminative ability of learned fMRI features.
\item
Our BMR is a general framework that can be applied to fMRI-based analysis of different brain disorders, such as autism spectrum disorder and HIV-associated neurocognitive disorder, 
as evidenced by its superior performance on ABIDE~\cite{di2014autism} and a private HAND dataset.
\end{itemize}

\section{Related Work}
\label{S2}

\subsection{Functional MRI Representation Learning}
Various machine learning methods have been used to learn latent representations of resting-state fMRI for brain disorder analysis~\cite{khosla2019machine}.
For example, 
Wee~\etal~\cite{wee2012constrained} proposed a constrained sparse linear regression model to estimate brain functional network (BFN) for mild cognitive impairment classification with resting-state fMRI.
Rosa~\etal~\cite{rosa2015sparse} designed a sparse
network-based predictive model that first constructed sparse inverse covariance matrices and then used a sparse support vector machine (SVM) for major depressive disorder detection. 
Gan~\etal~\cite{gan2021brain} proposed a multi-graph fusion method that
first integrated fully-connected BFNs and one nearest neighbor BFNs and then employed the L1SVM for brain disorder classification.
\if false
However, %
in most traditional machine learning approaches, the representation learning of BFN is not informed by downstream tasks but prior to classifier training. 

That is, the learned representation is fixed regardless of which downstream task is performed and thus model may achieve suboptimal performance.
\fi  
 However, these existing studies generally treat fMRI feature learning and downstream model training as two standalone steps, possibly leading to suboptimal performance due to heterogeneity between these steps.

Deep learning methods have been widely used for computer-aided brain disorder diagnosis with fMRI~\cite{yin2022deep}, by jointly conducting fMRI representation learning and downstream model training in an end-to-end manner. 
In particular, due to the graph structure nature of BFN, graph neural networks (GNNs) have shown significant superiority in fMRI representation learning.
For example, Ktena~\etal~\cite{ktena2018metric} proposed a 
siamese graph convolutional network (GCN) to estimate BFN for automated autism analysis.
Jiang~\etal~\cite{jiang2020hi} designed a hierarchical GCN framework for BFN embedding learning by efficiently integrating correlations among subjects in a population.
Hu~\etal~\cite{hu2022complementary} designed a complementary graph representation learning method to capture local and global patterns for fMRI-based brain disease analysis. 
Although these GNN-based methods can model spatial interactions among brain ROIs, they often neglect dynamic variations over time of fMRI data.
Considering that temporal dynamics conveyed in fMRI can provide discriminative information for brain disease diagnosis,
some GNN-based studies have paid more attention to spatiotemporally dynamic brain network analysis with fMRI data. 
For instance,  
Gadgil~\etal~\cite{gadgil2020spatio} introduced a novel spatiotemporal GCN, which captured temporal dynamics within fMRI series via 1D convolutional kernels, to learn dynamic grap representation for age and gender prediction. 
Even achieving promising results in fMRI representation learning, most of existing GNN models cannot explicitly preserve network topology of BFNs during dynamic graph learning. 

\subsection{Brain Functional Modularity Analysis}
From a graph-theoretic perspective, the BFN during the resting state exhibits a significant modular structure to facilitate efficient
information communication and cognitive function~\cite{meunier2010modular,sporns2016modular}. 
To better understand brain connectivity patterns, researchers have devoted considerable attention to analyzing brain modularity.
For example, Meunier~\etal~\cite{meunier2009age} explored age-related changes in brain modular organization and demonstrated significantly non-random modularity in young and older brain networks.
Arnemann~\etal~\cite{arnemann2015functional} tested the value of modularity metric to predict response to cognitive training after brain injury. 
Gallen~\etal~\cite{gallen2019brain} demonstrated that brain modularity could be regarded as a unifying biomarker of intervention-related plasticity by multiple independent studies.

Notably, previous neuroscience studies have demonstrated that there are three fundamental cognitive modules, \ie, salience network (SN), central executive network (CEN), and default mode network (DMN) in human brains. 
Specifically, SN mainly detects external stimuli and coordinates brain neural resources,
CEN performs high-level cognitive tasks (\eg, decision-making and rule-based problem-solving),
while DMN is responsible for self-related cognitive functions (\eg, mind-wandering and introspection)~\cite{goulden2014salience,menon2011large}.
These three modules have been consistently observed across different individuals and experimental paradigms~\cite{menon2011large,krishnadas2014resting}.
For example,
Menon~\etal~\cite{menon2011large} proposed a unifying triple network model comprised of CEN, DMN, and SN, providing a common framework for understanding behavioral and cognitive dysfunction across multiple brain disorders.  
Krishnadas~\etal~\cite{krishnadas2014resting} investigated disrupted resting-state functional connectivities within the triple network in patients with paranoid schizophrenia.  
Intuitively, such brain modularity structures can be employed as important prior knowledge to promote informative fMRI feature learning.  
However, existing studies typically fail to incorporate such important modularity prior into deep graph learning models for fMRI-based brain disorder analysis.

\if false
Recently, a few studies have recognized the importance of incorporating  brain modularity property as prior knowledge for BFN analysis.
For instance, based on a matrix-regularized network learning framework, Qiao~\etal~\cite{qiao2016estimating} encoded modularity prior into a machine learning model for fMRI-based brain disease analysis.
Notably, previous neuroscience studies have demonstrated that there are three fundamental cognitive modules (\ie, SN, CEN, and DMN) in human brain, 
where SN mainly detects external stimuli and coordinates brain neural resources,
CEN performs high-level cognitive tasks (\eg, decision-making and rule-based problem-solving),
and DMN is responsible for self-related cognitive functions (\eg, mind-wandering and introspection)~\cite{goulden2014salience,menon2011large}.
These three modules have been consistently observed across different individuals and experimental paradigms.
For example,
Menon~\etal~\cite{menon2011large} proposed a unifying triple network model comprised of CEN, DMN and SN, which provided a common framework for understanding behavioral and cognitive dysfunction across multiple disorders.  
Krishnadas~\etal~\cite{krishnadas2014resting} investigated disrupted resting-state functional connectivities within the triple network in paranoid schizophrenia.
However, to the best of our knowledge, existing studies typically fail to incorporate such important modularity prior into deep graph learning model for fMRI-based brain disease analysis.
%

\fi 




\section{Materials}
\label{S3}
\subsection{Data Acquisition}
A total of 534 subjects with rs-fMRI scans from the public Autism Brain Imaging Data Exchange (ABIDE) dataset~\cite{di2014autism} and a private  HIV-associated neurocognitive disorder (HAND) dataset are used in this work. 
On ABIDE, we identify patients with autism spectrum disorder (ASD) from healthy control (HC) subjects on the two largest sites (\ie, NYU and UM). 
Specifically, the NYU site includes $79$ ASDs and $105$ HCs, and the UM site includes $68$ ASDs and $77$ HCs.
On HAND, we perform two kinds of classification tasks, including (1) asymptomatic neurocognitive impairment with HIV (ANI) vs. HC classification,  and (2) intact cognition with HIV (ICH) vs. HC classification. 
Here, rs-fMRI data in the HAND are collected from a local hospital, including $67$ ANIs, $68$ ICHs and $70$ HCs.
The demographic information of the studied subjects from two datasets is reported in Table~\ref{demographic}.
\begin{table}[!tbp]
\setlength{\abovecaptionskip}{0pt}
\setlength{\belowcaptionskip}{0pt}
\setlength{\abovedisplayskip}{0pt}
\setlength{\belowdisplayskip}{0pt}
\renewcommand\arraystretch{1}
\centering
\caption{Demographic information of subjects from two largest sites (\ie, NYU and UM) of ABIDE dataset
 and a private HAND dataset. 
 ASD: autism spectrum disorder;  HC: healthy control; ANI: asymptomatic neurocognitive impairment with HIV; ICH: intact cognition with HIV; M: Male; F: Female; Std: Standard deviation.}
	\scriptsize
	\centering
	\setlength{\tabcolsep}{0.7pt}
	    \label{demographic}
		\begin{tabular*}{0.48\textwidth}{@{\extracolsep{\fill}} l|cccccc }
			\toprule
			~Dataset & Site & Category   &Subject~\# & Gender~(M/F) & Age~(Mean$\pm$Std) \\
		\midrule
			\multirow{4}{*} {~ABIDE}
	  	    &\multirow{2}{*}{NYU}   &ASD  &$79$	   &$68$/$11$	 &$14.52\pm6.97$ \\
			&\multirow{2}{*}        &HC   &$105$   &$79$/$26$	 &$15.81\pm6.25$ \\
		    \cline{2-6}
		    &\multirow{2}{*}{UM}    &ASD  &$68$    &$58$/$10$    &$13.13\pm2.41$ \\
			&\multirow{2}{*}     &HC   &$77$    &$59$/$18$   &$14.79\pm3.57$ \\
			\hline
			\multirow{3}{*} {~HAND}
		    &\multirow{3}{*}{--}   &ANI    &$67$	   &$67$/$0$	 &$33.07\pm6.18$ \\
		    &  &ICH  &$68$   &$68$/$0$   &$33.40\pm5.58$ \\
			&  &HC    &$70$   &$70$/$0$	 &$33.33\pm5.37$ \\
			\bottomrule
	    \end{tabular*}
\end{table}

\subsection{Data Preprocessing}
All rs-fMRI data from two datasets were preprocessed using the  Data Processing Assistant for Resting-State fMRI (DPARSF) pipeline~\cite{yan2010dparsf}. 
Specifically, for each fMRI, we first discarded the first 10 time points for magnetization equilibrium.
Then, we performed slice timing correction, head motion correction, and regression of nuisance covariates (\eg, white matter signals, ventricle, and head motion parameters). 
Afterward, the fMRI data were normalized into  montreal neurological institute (MNI) space, followed by spatial smoothing with a $4\,mm$ full width half maximum Gaussian kernel and band-pass filtering ($0.01-0.1\,Hz$). 
Finally, we extracted the mean rs-fMRI time series of $116$ ROIs defined by the automated anatomical labeling (AAL) atlas for each subject.  
\if false
Major steps include (1) magnetization equilibrium by trimming the first 10 volumes, 
(2) slice timing correction and head motion correction, 
(3) regression of nuisance covariates (\eg, white matter signals, ventricle, and head motion parameters), 
(4) spatial normalization to the MNI space, 
(5) bandpass filtering (0.01–0.10$\,$Hz). 
The average rs-fMRI time series of $116$ ROIs defined by the 
AAL atlas are extracted for each subject. 
\fi

\section{Proposed Methodology}


As illustrated in Fig.~\ref{pipeline}, the proposed BMR is comprised of three major components: (1) dynamic graph construction using sliding windows, (2) dynamic graph learning via a novel brain modularity-constrained graph neural network (MGNN), and (3) prediction and biomarker detection for interpretable brain disorder analysis, with details presented below. 







\subsection{Dynamic Graph Construction}
Brain functional network (BFN) derived from fMRI data can capture abnormal connectivity patterns caused by brain disorders by modeling complex dependencies among brain ROIs.
Given the fact that brain functional connectivity exhibits dynamic variability over a short period of time~\cite{gadgil2020spatio}, 
we construct a dynamic BFN using sliding windows for each subject~\cite{menon2019comparison}.
Denote BOLD signals obtained from rs-fMRI as {{$S\in \mathbb{R}^{N\times M}$}}, where {{$N$}} is the number of ROIs and  {{$M$}} is the number of time points. 
We first partition the fMRI time series into  {{$T$}} segments along the temporal dimension 
via overlapped sliding windows, with the window size of {{$\Gamma$}} and the step size of {{$\tau$}}. 
With each ROI treated as a specific node, we construct a fully-connected BFN by calculating Pearson correlation coefficients~\cite{schober2018correlation} between segmented fMRI time series of pairwise brain ROIs for each of $T$ segments, 
obtaining a set of symmetric matrices $\{X_{t}\}_{t=1}^{T}\in \mathbb{R}^{N\times N}$. 
Here, 
the original node feature for the $j$-th node is represented by the $j$-th row in $X_{t}$ for segment $t$ ($X_{t}$ is also called node feature matrix). 
Considering that a fully-connected BFN may 
contain some noisy or redundant information, following~\cite{kim2021learning}, we empirically retain the top $30\%$ (\ie, sparsity ratio) strongest edges in each FC network to generate an adjacent matrix  {{$A_{t}=(a_{ij})\in \{0,1\}^{N\times N}$ for the segment $t$, where $a_{ij}=1$ means there exists an edge between two nodes/ROIs and otherwise  $a_{ij}=0$.
Finally, the obtained dynamic graph sequence of each subject can be described as {{$G_{t}=\{A_{t}, X_{t}\}~ (t=1,\cdots,T)$}}.

\subsection{Dynamic Graph Representation Learning via MGNN}
As illustrated in the bottom of Fig.~\ref{pipeline}, with the constructed dynamic graph sequence {{$\{G_{t}\}_{t=1}^{T}$}}
as input, we design a brain modularity-constrained graph neural network (MGNN) for dynamic fMRI representation learning, including (1) spatial feature learning and (2) temporal feature learning,  which can 
simultaneously model spatial dependencies among brain ROIs and temporal dynamics over time. 
Notably, a novel \emph{brain modularity constraint} and a \emph{graph topology reconstruction constraint} are incorporated into MGNN to learn more interpretable and discriminative graph representations. 
\if false
More importantly, to learn more interpretable and discriminative graph representations, we add two unique constraints during spatial feature learning, including a \emph{modularity constraint} and a \emph{graph topology reconstruction constraint}. 
\fi

\subsubsection{Spatial Feature Learning}
\if false
Due to graph-structured property of BFN, graph neural networks (\eg, GAT~\cite{xu2018powerful} and GIN~\cite{xu2018powerful} and ),
which combine the advantage of graph theory and deep learning, have shown powerful potential to extract spatial features of fMRI data. 
Given the impressive representation ability of graph attention network (GAT) that can assign adaptive weights between nodes/ROIs,
we employ GAT as a spatial encoder to effectively mine topological information of BFN in this work.
\fi 

Considering the graph-structured property of BFNs, we employ a graph attention network (GAT) as the spatial feature encoder to model spatial dynamic representation of BFNs in this work. 
Taking the segment $t$ as an example, 
the spatial encoder takes the node feature matrix $X_{t}=[X_{1}^{t},X_{2}^{t},\cdots,X_{N}^{t}]^{\top}$ ($X_{i}^{t}\in \mathbb{R}^{1\times N}$) and the graph adjacent/topology matrix $A_{t}$ as input. 
Denote $\mathcal{N}_i^t$ as the neighboring node set of the $i$-th node and $\oplus$ as the concatenation operation. 
The to-be-learned connection weight (also called spatial attention coefficient) between the $i$-th ROI and its neighborhood ROI $j$ can be formulated as: 
\begin{equation}
\small
\begin{aligned}
\alpha^{t}_{ij}=\frac{{exp}\left(\psi ([X_{i}^{t}W^{t} \oplus X_{j}^{t}W^{t}]\eta^{t} ) \right)}{\sum\nolimits_{v\in \mathcal{N}_{i}^{t}} {exp} \left(\psi([X_{i}^{t}W^{t} \oplus X_{v}^{t}W^{t}]\eta^{t})\right)},
\\
\end{aligned}
\label{EQ1}
\end{equation}
where $\psi$ is a nonlinear activation function (\ie, LeakyRelu), 
$W^{t}\in \mathbb{R}^{N\times N'}$ is a shared  transformation matrix that maps the original $N$-dimensional node feature vector to an $N'$-dimensional vector, and 
${{\eta^{t}}\in \mathbb{R}^{2N'}} $ is a to-be-learned weight vector.  
Then, the updated node representation is expressed as: 
\begin{equation}
\small
\begin{aligned}
F_{i}^{t} =\sum\nolimits_{j\in \mathcal{N}_{i}^{t}}\alpha^{t}_{ij}X_{j}^{t}W^{t},
\\
\end{aligned}
\label{Eq1}
\vspace{-2pt}
\end{equation}
where $F_{i}^{t}\in \mathbb{R}^{1\times N'}$ is the new embedding of the node $i$ after aggregating neighboring node representations.
To model different types of spatial dependencies/relationships among ROIs/nodes,
we employ a multi-head attention mechanism, which first calculates node representation using multiple attention heads in parallel and then averages them.
Mathematically, the output feature $H_{i}^{t}\in \mathbb{R}^{1 \times N'}$ of the node $i$ generated by the multi-head attention mechanism can be written as follows: 
\begin{equation}
\small
\begin{aligned}
H_{i}^{t} =\sigma(\frac{1}{K}\sum\nolimits_{k=1}^{K}{F_{i}^{t,k}}),
\end{aligned}
\label{Eq1}
\end{equation}where $\sigma$ is a nonlinear function and $K$ is the number of attention heads.
Given $N$ nodes, the new node-level embedding of BFN can be expressed as $H^{t}=[H_{1}^{t},\cdots,H_{N}^{t}]^{\top}$ at segment $t$. 

\begin{figure}[!t]
\centering
\setlength{\abovecaptionskip}{0pt}
\setlength{\belowcaptionskip}{-2pt}
\setlength{\abovedisplayskip}{-2pt}
\setlength{\belowdisplayskip}{-2pt}
\includegraphics[width=0.48\textwidth]{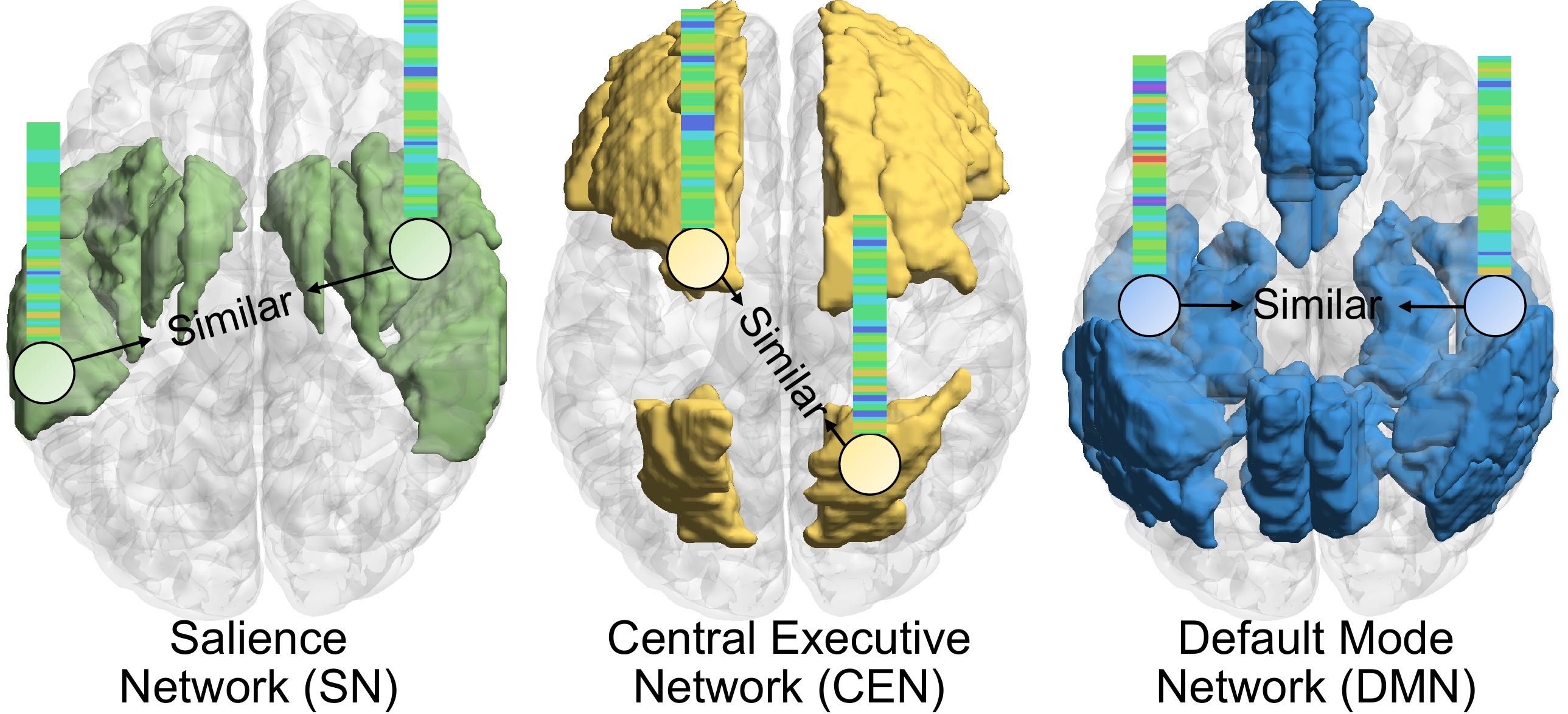}
\caption{Illustration of the proposed modularity constraint with three fundamental cognitive modules, \ie, salience network (SN), central executive network (CEN), and default mode network (DMN), where nodes within the same module are encouraged to share similar representation.
}
\label{modules}
\end{figure}
\emph{(1) \textbf{Brain Modularity Constraint.}}
As an important property of BFN~\cite{gallen2019brain}, modularity provides valuable insights into the organization and integration of brain networks.
In general, each module is comprised of densely interconnected brain ROIs that are sparsely connected to ROIs in other modules~\cite{sporns2016modular}. 
Due to the high clustering of connections between ROIs within the module, 
the brain can locally process specialized cognitive function (\ie, episodic memory processing) with low wiring cost~\cite{meunier2010modular}.
Previous studies have demonstrated that SN, CEN, and DMN are three fundamental neurocognitive modules in human brains.
Based on such prior knowledge, we reasonably assume that \emph{representations of nodes within the same neurocognitive module tend to be similar to each other}.

Accordingly, we design a unique \emph{brain modularity constraint} during spatial fMRI feature learning in BMR.  
As illustrated in Fig.~\ref{modules}, the modularity constraint explicitly encourages the nodes belonging to the same module to share similar features. 
Based on Cosine distance metric, this constraint can be mathematically formulated as follows: 
\if false
Based on such prior knowledge, we reasonably assume that \emph{to-be-learned representations of nodes within the same neurocognitive module tend to be similar to each other}, with detailed illustration shown in Fig.~\ref{modules}. 
In this work, we employ the cosine similarity metric to evaluate the similarity between two node-level features. 
Mathematically, the proposed \emph{modularity constraint} can be formulated as:
\fi 
\begin{equation}
\small
L_{M}
= -\sum\nolimits_{t=1}^{T}\sum\nolimits_{c=1}^{C}\sum\nolimits_{i,j=1}^{N_c}  \frac{H_{i}^{t,c}\cdot H_{j}^{t,c}}{\Vert H_{i}^{t,c}\Vert \cdot \Vert H_{j}^{t,c}\Vert}, 
\label{module constraint}
\end{equation} where {{$H_{i}^{t,c}$}} and {{$H_{j}^{t,c}$}}
are representations of node $i$ and node $j$ within the $c$-th module (with $N_c$ ROIs) at segment $t$, and {{$C$}} is the number of modules ($C=3$ in this work). 
With Eq.~\eqref{module constraint}, 
we encourage the BMR to focus on brain intrinsic modular organization during fMRI representation learning, thus enhancing discriminative power of learned fMRI features.
\if false
It has been demonstrated that the central executive network (CEN), salience network (SN) and default mode network (DMN) are three crucial neurocognitive modules in the brain, 
where CEN performs high-level cognitive tasks (\eg, decision-making and rule-based problem-solving),
SN mainly detects external stimuli and coordinates brain neural resources,
and DMN is responsible for self-related cognitive functions~\cite{goulden2014salience,sporns2016modular,bertolero2015modular}. 
The ROIs/nodes within a module are densely inter-connected, resulting in a high degree of clustering between nodes from the same module. 
Based on such prior knowledge, we reasonably assume that the \emph{learned embeddings of nodes within the same neurocognitive module tend to be similar}.
We develop a novel modularity constraint to encourage similarity between paired node-level embeddings in the same module. 
Mathematically, the proposed \emph{modularity constraint} 
is formulated as:
\begin{equation}
L_{M}
= -\sum\nolimits_{t=1}^{T}\sum\nolimits_{k=1}^{K}\sum\nolimits_{i,j=1}^{N_k}  \frac{h_{i}^{t,k}\cdot h_{j}^{t,k}}{\Vert h_{i}^{t,k}\Vert \cdot \Vert h_{j}^{t,k}\Vert} 
\label{Eq2}
\end{equation} where {{$h_{i}^{t,k}$}} and {{$h_{j}^{t,k}$}}
are embeddings of two nodes in the $k$-th module (with $N_k$ ROIs) at segment $t$, and {{$K$}} is the number of modules ($K=3$ in this work). 
With Eq.~\eqref{Eq2}, 
we encourage the MSGNN to focus on modular brain structures during representation learning, thus improving discriminative ability of fMRI features. 

\fi

\emph{(2) \textbf{Graph Topology Reconstruction Constraint.}}
To further improve discriminative ability of learned representations, we propose to preserve original topology information of input BFNs by reconstructing the adjacent matrix $A_{t}$ at segment $t$ ($t=1,\cdots,T$). 
Motivated by variational graph autoencoder~\cite{kipf2016variational}, we design a graph decoder in BMR to first predict the presence of edges between pairwise nodes based on the inner-product of two latent node representations, yielding a reconstructed adjacent matrix {{$\Hat{A}_{t}=\sigma(H_{t}\cdot H_{t}^{\top})$}} for segment $t$, where {{$\sigma$}} is a nonlinear mapping function. 
We then propose a 
\emph{graph topology reconstruction constraint} to preserve network topology of input BFNs, formulated as: 
\begin{equation}
\small
L_{R}
= \sum\nolimits_{t=1}^{T}\xi({A}_{t},\Hat{A}_{t}), 
\label{reconstruct}
\end{equation}where {{$\xi$}} is a cross-entropy loss function. 
With Eq.~\ref{reconstruct}, the reconstructed graph is encouraged to closely resemble the original graph as far as possible, so that the learned node embeddings can effectively capture underlying FCN structure and relationships among ROIs.

To generate graph-level  representations, 
we further apply a squeeze-excitation readout operation~\cite{hu2018squeeze} based on learned node-level representations.
For segment $t$, the graph-level spatial representation 
is calculated as:
\begin{equation}
y_{t}=H_{t}\Phi \left(P^{2}\sigma(P^{1}H_{t}\phi_{mean})\right),
\label{Eq3}
\end{equation}
where {{$\Phi$}} is a sigmoid function, {{$P^{1}$ and $P^{2}$}} are learnable weight matrices, 
and {{$\phi_{mean}$}} denotes the average operation. 

\subsubsection{Temporal Feature Learning}
As shown in the bottom right of Fig.~\ref{pipeline}, to further capture temporal dynamics within fMRI series, a single-head transformer encoder is employed to effectively model long-range dependencies across different segments.
Here, temporal attention can be measured by a self-attention mechanism in the transformer.
Especially, with spatial graph representations $Y=[y_{1},\cdots,y_{T}]$ as input,
temporal attention weights can be described as:
\begin{equation}
Z=Softmax(\frac{QK^{\top}}{\sqrt{d}}),
\small
\label{Eq3}
\end{equation}where $Q=\phi_{1}(Y)$, $K=\phi_{2}(Y)$, $\phi_{1}$ and $\phi_{2}$ are two linear transformations, $d$ is a scaling factor to stabilize attention mechanism.
Thus, the graph representation with spatiotemporal attention can be expressed as $Y'=\Psi [Z\phi_{3}(Y)]$, where $Y'=[y_{1}',\cdots,y_{T}']$, $\phi_{3}$ is a linear transformation, and  $\Psi$ represents the feed-forward network for further feature abstraction.
After that, 
we sum the updated graph representation sequence {{$\{{y}_{i}'\}_{i=1}^{T}$}} to obtain the final whole-graph embedding for subsequent brain disorder analysis.

\begin{table*}[!t]
\setlength{\abovecaptionskip}{0pt}
\setlength{\belowcaptionskip}{0pt}
\setlength{\abovedisplayskip}{0pt}
\setlength{\belowdisplayskip}{0pt}
\scriptsize
\textbf{\renewcommand{\arraystretch}{0.9}}
	\caption{Classification results  of competing methods and the proposed BMR on the two largest sites of ABIDE.
	Results are shown in the form of ``mean(standard deviation)'' and  the best results are shown in bold.
 The term `*' represents the proposed BMR is statistically significantly different from a competing method.
	}
	\label{ABIDE_main}
	\centering
	\setlength{\tabcolsep}{0.5pt}
    \begin{tabular*}{1\textwidth}{@{\extracolsep{\fill}}l|cccccc c cccccc}\toprule
    \multirow{2}{*}{Method} & \multicolumn{6}{c}{ASD vs. HC classification on NYU} && \multicolumn{6}{c}{ASD vs. HC classification on UM} \\
\cmidrule{2-7} \cmidrule{9-14}
		 & AUC (\%) & ACC (\%) & F1 (\%) & SEN (\%) & SPE (\%) & BAC (\%) &~& AUC (\%) & ACC (\%) & F1 (\%) & SEN (\%) & SPE (\%) & BAC (\%) \\
		\midrule
        SVM&56.64(2.89)$^*$ &	54.83(3.11)	& 48.64(3.38)&51.48(4.56)	&57.88(4.66) &	54.69(3.09)& &
      53.61(4.25)$^*$&53.32(3.63)&49.32(4.20)&50.29(4.53)&	56.60(4.66)	&53.45(3.76)\\
        XGBoost&61.95(0.56)$^*$&63.00(1.63)&51.49(1.84)	&47.99(2.68)&\textbf{75.91(3.72)}&61.95(0.56)& &
58.81(0.83)$^*$&58.62(1.89)&50.19(1.95)&47.61(3.64)	&70.01(3.94)&58.81(0.84)\\
        Random Forest &61.21(3.09)$^*$&61.10(4.48)&49.31(4.62)&46.04(4.12)&74.06(5.23)&60.05(4.67)
        & &      57.25(2.77)$^*$&56.13(3.66)&49.25(4.87)&47.73(5.83)&65.68(4.11)&56.70(4.04)
        \\
        MLP&58.64(2.51)$^*$&58.28(1.60)&46.66(4.39)&45.22(7.22)&68.24(5.48)&56.73(1.64)&&
        58.17(2.89)$^*$&54.48(3.00)&48.06(1.81)&47.67(2.47)&62.70(5.99)&55.19(2.78)
        \\
        GCN&67.53(3.34)&63.59(3.05)&53.63(3.75) &50.99(5.09)&73.54(4.69)&62.26(2.82)& &
    66.74(2.58)$^*$&60.00(2.96)&55.30(3.07)&54.61(4.42)&	66.50(5.22)&60.56(2.57)\\
        
        GIN&61.43(3.45)$^*$&57.04(2.48)&48.24(2.13)&49.35(5.39)&64.94(5.70)&57.14(1.16)&&
        58.94(3.19)$^*$&56.89(2.73)&50.47(2.73)&49.62(3.23)&64.92(3.41)&57.27(2.40)
        \\
        GAT&64.87(2.64)$^*$&60.12(2.64)&52.14(3.83)	&52.96(4.92)&66.12(3.15)&59.54(2.79)& &
        67.34(3.26)$^*$&60.90(3.58)&55.54(4.59)&54.98(5.39)&	68.21(5.15)&61.60(2.95)\\
        BrainGNN&66.89(2.90)$^*$&63.21(3.15)&56.31(4.17)&	57.12(4.81)&68.51(3.03)&62.82(3.24)& &
       65.91(2.47)&62.69(2.55)&57.18(1.21)&55.47(3.25)	&68.09(5.67)&61.78(2.06) \\
        STGCN & 66.69(0.87)$^*$&61.54(1.65)&45.55(2.23)&53.64(2.34)&68.42(1.75)&61.03(1.53)& 
        &64.01(0.14)$^*$&63.90(0.10)&44.19(3.91)& 55.95(1.18)& \textbf{72.16(0.56)}& 64.07(0.14)
        \\
       BMR (Ours) &\textbf{73.24(4.48)}&\textbf{67.10(4.29)}&\textbf{62.16(4.53)}&\textbf{64.74(5.08)}&69.39(6.22)&\textbf{67.06(4.38)}&&
        \textbf{70.55(5.22)}&\textbf{65.28(2.37)}&\textbf{62.25(2.14)}&\textbf{63.21(2.71)}&67.42(4.79)&\textbf{65.31(3.08)}\\
		\bottomrule
	\end{tabular*}
\end{table*}

\begin{table*}[!t]
\setlength{\abovecaptionskip}{0pt}
\setlength{\belowcaptionskip}{0pt}
\setlength{\abovedisplayskip}{0pt}
\setlength{\belowdisplayskip}{0pt}
\scriptsize
\textbf{\renewcommand{\arraystretch}{0.9}}
	\caption{Classification results  of competing methods and the proposed BMR on HAND.
	Results are shown in the form of ``mean(standard deviation)'' and  the best results are shown in bold.
 The term `*' represents the proposed BMR is statistically significantly different from a competing method.
	}
	\label{HAND_main}
	\centering
	\setlength{\tabcolsep}{0.5pt}
    \begin{tabular*}{1\textwidth}{@{\extracolsep{\fill}}l|cccccc c cccccc}\toprule
	\multirow{2}{*}{Method} & \multicolumn{6}{c}{ANI vs. HC classification on HAND} && \multicolumn{6}{c}{ICH vs. HC classification on HAND}  \\
 \cmidrule(l){2-7} \cmidrule(l){9-14}
		 & AUC (\%) & ACC (\%) & F1 (\%) & SEN (\%) & SPE (\%) & BAC (\%) && AUC (\%) & ACC (\%) & F1 (\%) & SEN (\%) & SPE (\%) & BAC (\%) \\
		\midrule

        SVM&61.73(3.21)$^*$&57.28(2.68)&56.25(2.98)	&58.25(4.27)&57.22(2.87)&57.73(2.72)&&
        53.91(3.54)$^*$&52.52(4.75)&51.86(5.61)&53.95(6.29)&51.50(7.25)&52.73(5.14) \\        XGBoost&56.72(3.59)$^*$&53.34(2.83)&51.11(2.82)&51.70(3.61)&56.70(6.59)&54.21(2.76)&&
        55.05(3.66)$^*$&53.04(2.19)&51.29(2.27)&52.34(3.46)&55.55(3.51)&53.95(1.92)\\
        Random Forest&64.64(1.46)&58.61(1.54)&57.88(3.03)&60.49(5.13)&59.51(5.45)&60.00(1.52)&&
        57.11(4.23)$^*$&53.10(4.79)&51.70(3.95)&53.31(4.81)&55.89(8.04)&54.60(4.70)
        \\
        MLP&64.68(3.82)$^*$&58.97(2.69)&57.37(2.49)&59.05(2.82)&59.55(3.85)&59.30(2.28)&&
        56.44(4.02)&55.25(3.40)&55.81(2.55)&59.30(2.09)&52.24(6.86)&55.77(3.40)\\
        GCN&64.23(2.13)$^*$&59.60(2.32)&57.78(3.29)&	58.65(5.06)&61.65(4.89)&60.15(2.39)&&
        55.42(6.44)$^*$&54.65(1.49)&53.08(5.09)&55.60(9.52)&53.63(10.20)&54.62(1.60)\\
        GIN&65.93(3.61)$^*$&60.34(2.25)&58.40(3.06)&59.00(4.57)&62.36(3.49)&60.68(2.72)&&
        57.78(3.60)$^*$&54.80(3.69)&53.78(4.11)&55.90(4.58)&56.39(4.04)&56.14(3.21)\\
        GAT&65.72(3.86)$^*$&61.11(2.99)&59.24(0.78)&	\textbf{62.53(5.83)}&59.47(6.00)&61.00(0.09)&&
        54.42(3.01)$^*$&53.67(4.03)&54.41(4.00)&60.06(8.78)&50.12(7.77)&55.09(4.57)\\
        BrainGNN&63.04(3.23)$^*$&60.26(2.60)&	56.87(3.78)&56.32(5.95)&63.70(5.59)&	60.01(1.78)&&
        58.14(4.75)&56.74(2.23)&\textbf{57.38(2.84)}&\textbf{61.48(3.80)}&51.20(3.45)&56.34(2.36)\\
        STGCN & 53.81(2.94)$^*$&51.20(4.00)&52.45(3.78)&57.40(5.45)&47.40(5.64)&52.40(4.02)&&
        55.22(8.36)$^*$&51.11(7.62)&50.88(10.36)&55.34(22.93)&49.48(8.16)&52.41(8.40)\\
       BMR (Ours) &\textbf{67.91(3.49)}&\textbf{64.03(4.97)}&\textbf{61.78(6.89)}&62.39(8.68)&\textbf{66.25(3.81)}&\textbf{64.32(5.16)}&&
        \textbf{60.26(4.98)}&\textbf{57.49(4.15)}&55.15(5.58)&54.82(7.26)&\textbf{61.12(6.94)}&\textbf{57.97(4.34)}\\
		\bottomrule
	\end{tabular*}
\end{table*}

\subsection{Prediction and Interpretable Biomarker Detection}
The whole-graph embedding is then fed into two fully connected layer and a Softmax layer for brain disease prediction. 
The objective function of our BMR can be formulated as:
\begin{equation}
\small
L=L_{C}+\lambda_{1}L_{R}+\lambda_{2}L_{M},
\label{loss}
\end{equation} where {{$L_{C}$}} is a cross-entropy loss for prediction, $L_{R}$ and $L_{M}$ denote the proposed graph topology reconstruction constraint and brain modularity constraint, respectively, 
while {{$\lambda_{1}$}} and {{$\lambda_{2}$}} are two hyperparameters. 

\if false
\begin{equation}
\small
L_{C} = -[q  log(p)+(1-q) log(1-p)],
\label{Eq9}
\end{equation}where $p$ and $q$ denote the predicted result and true label (\ie, $0$ or $1$).
\fi 

\if false
--\emph{\textbf{Overall Learning Objective}}
Assuming that a dataset contains $Q$ training subjects,
the final loss function of the BMR is defined as:
\begin{equation}
L=\sum\nolimits_{q=1}^{Q}L_{C}^{q}+\lambda_{1}L_{R}^{q}+\lambda_{2}L_{M}^{q}
\label{Eq5}
\end{equation}
To introduce final loss function

-- \emph{\textbf{Automatically Identified Biomarker}}
\fi

To facilitate interpretation of our learned graph representations, we further analyze spatial attention among brain ROIs, which can provide potential biomarkers for brain disorder diagnosis.
Specifically, based on spatial attention coefficients described in Eq.~\ref{EQ1}, we first obtain spatial attention matrices of $T$ segments and average them to generate a spatial attention matrix for each subject.
Then, we take the upper triangle elements of each attention matrix, resulting in a $6,670$-dimensional vector.
Finally, we employ $t$-test to select discriminative features by calculating group differences between patients and healthy controls, and also map these features to their original brain space to detect the most discriminative functional connectivities in brain disease detection. 
The specific biomarker analysis 
 will be introduced in Section~\ref{biomarkerAnalysis}

\subsection{Implementation Details}\label{subsec_implementation}
The proposed BMR is implemented in PyTorch using a single GPU (NVIDIA TITAN Xp with 12GB memory). 
The Adam optimizer is used for optimization, 
with the learning rate of {{$0.0001$}}, training epochs of {{$40$}}, batch size of {{$8$}}, window size of $\Gamma=40$ and step size of $\tau=20$. 
Within the $c$-th module (with $N_{c}$ ROIs), we randomly select $m=50\%$ of all {{$\frac{N_{c}(N_{c}-1)}{2}$}} paired ROIs to constrain the BMR. 
The hyperparameters (\ie, $\lambda_{1}$ and $\lambda_{2}$) in Eq.~\eqref{loss} are determined via a cross-validation strategy (see Section~\ref{ExperimentSetting} and Section~\ref{parameter selection}).


\section{Experiments}


\subsection{Experimental Settings}
\label{ExperimentSetting}
A 5-fold cross-validation (CV) strategy is employed in the experiments. 
Besides, within each fold, we randomly select 20\% of training samples as the validation set to determine the optimal parameters.
We repeat above 5-fold CV process five times
to avoid bias caused by data partition and record the mean and standard deviation results. 
Six metrics are used to evaluate classification performance, including the area under the receiver operating characteristic curve (AUC), classification accuracy (ACC), F1 score (F1), sensitivity (SEN), specificity (SPE), and balanced accuracy (BAC). 
Paired sample $t$-test is used to perform statistical significance analysis
between the BMR and each of competing methods.

\begin{figure*}[t]
\setlength{\abovecaptionskip}{0pt}
\setlength{\belowcaptionskip}{-2pt}
\setlength{\abovedisplayskip}{-2pt}
\setlength{\belowdisplayskip}{-2pt}
\centering
\includegraphics[width=0.96\textwidth]{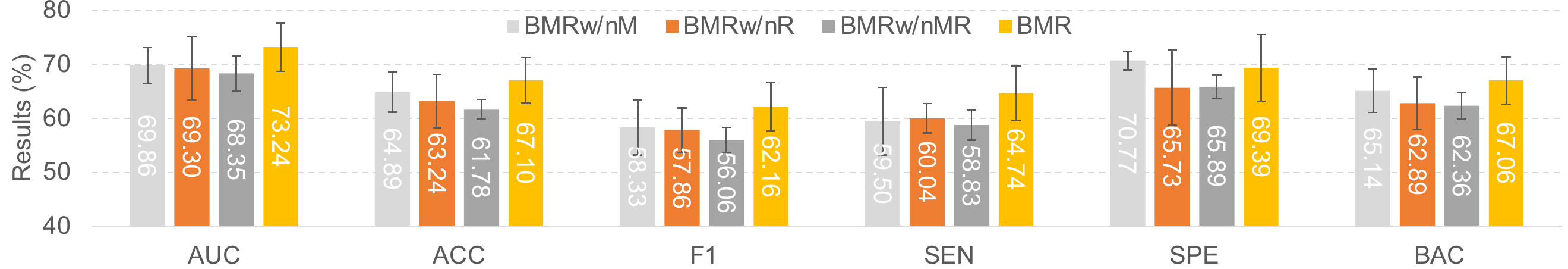}
\caption{Performance of the BMR and its three variants in ASD vs. HC classification on NYU site of ABIDE dataset.}
\label{ablation}
\end{figure*}

\subsection{Competing Methods}\label{sec_competing_methods}
We compare the proposed BMR with three conventional machine learning methods: SVM~\cite{noble2006support}, XGBoost~\cite{chen2016xgboost}, and Random Forest~\cite{biau2016random}, as well as six deep learning methods: multilayer perceptron (MLP)~\cite{murtagh1991multilayer}, GCN~\cite{kipf2016semi}, graph isomorphism network (GIN)~\cite{kim2020understanding}, graph attention network (GAT)~\cite{velivckovic2017graph}, BrainGNN~\cite{li2021braingnn}, and spatio-temporal graph convolutional network (STGCN)~\cite{gadgil2020spatio}, with details introduced below. 

(1) \textbf{SVM}:
In this method, we first construct a BFN based on rs-fMRI for each subject by measuring Pearson correlation (PC) coefficients between pairwise brain ROIs.
Then, we extract multiple node statistics (\ie, degree centrality, clustering coefficient, betweenness centrality, and eigenvector centrality) of each BFN and concatenate them into a $464$-dimensional vector. 
Finally, the vectorized fMRI representation is fed into a linear SVM (with default parameter $C=1$) for classification. 

(2) \textbf{XGBoost}:
Similar to SVM, we first construct a BFN based on PC for each subject and then  concatenate the same node statistics into a vectorized representation, followed by XGBoost (with default parameters) for classification. 

(3) \textbf{Random Forest}:
This method uses the same fMRI features as SVM and XGBoost, followed by a random forest classifier (with default parameters).  

(4) \textbf{MLP}:
Similar to above three methods,
we first extract node features to represent each subject, and then use two fully connected layers 
for feature abstraction  and a Softmax layer for brain disorder prediction. 

(5) \textbf{GCN}:
In this method, we first construct a BFN using PC for each subject.
Then, two graph convolutional layers are used to update and aggregate node-level representations. 
Finally, graph-level representations are generated via readout operation based on node representations,
followed by two fully connected layers for classification.  

(6) \textbf{GIN}:
Similar to GCN, 
this method first constructs a BFN for each subject
and then uses two GIN layers that implement Weisfeiler-Lehman graph isomorphism test in a neural network for feature learning.
Finally, we obtain graph representations via a readout operation, followed by two fully connected layers and a Softmax layer for classification.

(7) \textbf{GAT}:
Different from GCN and GIN, the GAT uses a graph attention mechanism to learn adaptive edge weights between brain ROIs.
In this method, we use two graph attention layers to learn spatial features, a readout operation to generate graph-level vectors, two fully connected layers, and a Softmax layer for classification. 
Similar to our BMR, the number of attention heads is $4$ for each graph attention layer. 

(8) \textbf{BrainGNN}: 
BrainGNN is a popular GNN model specially designed for fMRI analysis, containing an ROI-selection layer for highlighting salient brain ROIs.
With constructed BFNs as input, BrainGNN uses two ROI-aware graph convolutional layers to learn node embeddings, followed by ROI pooling layers to reduce the size of graph. 
Then, a readout operation is used to convert node-level features into graph-level representations,
followed by two fully connected layers and a Softmax layer for classification.

(9) \textbf{STGCN}:
The STGCN  can jointly capture spatial and temporal information of FCNs via spatiotemporal graph convolution (ST-GC) units.
In this method,
we first employ two ST-GC units for learning spatiotemporal features  with BOLD signals of ROIs as input.
Then, dynamic graph representations are generated via readout operation, followed by two fully connected layers for disease prediction.

For a fair comparison, we use the same number of hidden layers (\ie, 2) and the same number of neurons in each hidden layer (\ie, 64) for five GNN-based methods (\ie, GCN, GIN, GAT, BrainGNN, and STGCN). 
Additionally, we use the same input BFN data for four GNN-based methods (\ie, GCN, GIN, GAT, and Brain GNN), while the STGCN employs the processed fMRI time series as input.


\if false
To evaluate model performance, we compare the BMR with three conventional machine learning  methods, including 1) linear \textbf{SVM} that 
first concatenates  multiple node statistics (\ie, degree centrality, clustering coefficient, betweenness centrality and eigenvector centrality) of each BFN into a $464$-dimensional vector and feeds it into an SVM classifier.
2) \textbf{XGBoost} 
that feeds the above node-level features (similar to SVM) into an XGBoost classifier.
(3) \textbf{Random Forest}~\cite{liaw2002classification} that employs random forest for classification with the same input features as SVM,
and six deep learning methods, including: 
(4) \textbf{MLP} that takes node statistics (similar to SVM) as input and uses two fully connected layers (with $20$ hidden neurons) and a Softmax layer for further feature learning and classification. 
(5) \textbf{GCN}~\cite{kipf2016semi} that models brain spatial patterns via graph convolutional layers,
(6) \textbf{GIN}~\cite{kim2020understanding} that generalizes the Weisfeiler-Lehman (WL) graph isomorphism test 
and shows effective graph representation ability,
(7) \textbf{GAT}~\cite{velivckovic2017graph}  that uses an attention mechanism to learn adaptive edge weights between nodes/ROIs, 
(8) \textbf{BrainGNN}~\cite{li2021braingnn} that maps regional and cross-regional activation patterns of fMRI for brain disorder analysis. 
and (9) \textbf{STGCN}~\cite{gadgil2020spatio} that can jointly capture spatial and temporal information of FCNs via spatiotemporal graph convolution (ST-GC) units.
Note that the settings of network layers and hidden neurons in GNN-based methods (\ie, GCN, GIN, GAT, BrainGNN and STGCN) are same as the BMR.
\fi




 


\subsection{Classification Results}
The quantitative results of the proposed BMR and nine competing methods on ABIDE and HAND are reported in Table~\ref{ABIDE_main} and Table~\ref{HAND_main}, respectively, 
where `*' represents the proposed BMR is statistically significantly different from a specific competing method via paired sample $t$-test.
From Tables~\ref{ABIDE_main}-\ref{HAND_main}, we have the following interesting observations.

\emph{First}, our BMR is superior to traditional machine learning methods (\ie, SVM, XGBoost, and Random Forest) by a significant margin on two datasets (\ie, ABIDE and HAND).
For example, in terms of AUC values, the BMR yields the improvement of $16.6\%$, $11.29\%$, $12.03\%$ compared with SVM, XGBoost, Random Forest on the NYU site of ABIDE, respectively. 
The possible reason is that our BMR can learn informative fMRI representation in an end-to-end manner as needed for downstream tasks compared with these traditional methods that rely on handcrafted node features. 
\if false
\emph{Second}, 
in most cases, 
the BMR outperforms the MLP method without modeling spatial information between brain ROIs, 
which further validates the superiority of GNN-based model in graph/BFN representation learning. 
\fi 
\emph{Second}, compared with six  deep models (\ie, MLP, GCN, GIN, GAT, BrainGNN, and STGCN), our BMR achieves better performance in terms of most metrics on two datasets.
For instance, in the task of ANI vs. HC classification on HAND, the BMR improves the AUC value by $4.87\%$, compared with BrainGNN (a GNN-based model specially designed for brain network analysis). 
This is probably because our BMR not only focuses more on three inherent functional modules in the brain but also preserves the original topology structure during graph learning, 
resulting in more discriminative fMRI representation for classification. 
\emph{Furthermore}, it can be seen the BMR consistently outperforms STGCN that models short-range temporal dynamics within fMRI via a convolution operation. 
The possible reason is that our BMR can not only capture long-range temporal dependencies within fMRI series via a transformer encoder, but also incorporate crucial modularity prior into the process of dynamic graph representation learning, thereby achieving better classification performance. 
\if false
Different from previous dynamic graph representation learning model (\ie, STGCN), our BMR can not only capture long-range temporal dependencies within fMRI series via transformer encoder, but also  incorporate crucial modularity prior into the process of dynamic graph representation learning, thereby
achieving better classification performance.
\fi 

\if false

{\color{blue}On the other hand, Reasons for minor improvement on MDD. XXXXXXX}
\fi


\if false 
It can be seen that our MDRL generally outperforms two shallow methods (\ie, SVM and XGBoost) that rely on handcrafted node features without modeling whole-graph topological information. 
Compared with 4 SOTA deep learning methods, 
our MDRL achieves superior performance in terms of most metrics in three tasks. 
For instance, for ASD vs. HC classification on NYU of ABIDE (see Table~\ref{mainABIDE}), the AUC value of MDRL is improved by $5.7\%$ compared with BrainGNN (a SOTA method designed for brain network analysis). 
This implies the MDRL 
can learn discriminative graph representations to boost fMRI-based 
learning performance. 
\fi

\if false
\begin{table}[!tbp]
\setlength{\abovecaptionskip}{-0pt}
\setlength{\belowcaptionskip}{-2pt}
\setlength{\abovedisplayskip}{-2pt}
\setlength{\belowdisplayskip}{-2pt}
	\scriptsize
	\setlength{\tabcolsep}{1}
    \renewcommand{\arraystretch}{0.9}
	\caption{Classification results  of baseline methods and the proposed BMR on HAND.
	Best results are shown in bold and the term `*' represents the proposed BMR is statistically significantly different from a baseline method.}
	\label{HAND_main}
	\centering
	\begin{tabular*}{0.48\textwidth}{@{\extracolsep{\fill}} l|cccccc}
		\toprule
        ~Method  & AUC (\%) & ACC (\%) & F1 (\%) & SEN (\%) & SPE (\%) & PRE (\%)   \\
        \midrule
              SVM$^*$&50.67±3.83&50.67±3.83&50.67±3.83&50.67±3.83&50.67±3.83&50.67±3.83  \\
        XGBoost\\
        Random Forest\\
        MLP\\
        wck-CNN\\
        GCN\\
        GIN\\
        GAT\\
        BrainGNN\\
        STGCN\\
       BMR (Ours) &\\
		\bottomrule
	\end{tabular*}
\end{table}

\fi

\subsection{Ablation Study}
To investigate the effectiveness of key components in the proposed method, we compare the BMR with its three variants: 
(1) \textbf{BMRw/oM} without the modularity constraint,
(2) \textbf{BMRw/oR} without the graph topology reconstruction constraint,
and (3) \textbf{BMRw/oMR} that only uses GAT and Transformer layers for spatiotemporal representation learning, without the two constraints.
The experimental results yielded by these four methods in ASD vs. HC classification on NYU from ABIDE are reported in Fig.~\ref{ablation}.

From Fig.~\ref{ablation}, we can see that BMR outperforms BMRw/oM without considering the inherent modular structure in the brain. 
This implies that incorporating brain modularity prior to fMRI representation can help promote  classification performance by learning more discriminative features. 
Besides, the BMR is superior to BMRw/oR without performing graph topology reconstruction during fMRI representation learning. 
The underlying reason is the proposed graph topology reconstruction constraint helps capture intrinsic spatial information among brain ROIs. 
In addition, the BMRw/oMR without the proposed two constraints achieves the worst performance in most cases compared with its three counterparts (\ie, BMRw/oM, BMRw/oR, and BMR), which further validates the necessity of including modularity constraint and graph topology constraint. 



\section{Discussion}\label{sec_discussion}

\subsection{Discriminative Brain ROI and Functional Connectivity}\label{biomarkerAnalysis}
We also visualize the top 10 discriminative functional connectivities (FCs) identified by the proposed BMR on different datasets (\ie, ABIDE and HAND) in Fig.~\ref{fig_FCs}.
Note that the thickness of each line represents discriminative ability of the corresponding FC (inversely proportional to the $p$-value obtained by $t$-test).
For ASD identification (see Fig.~\ref{fig_FCs}~(a)), the most discriminative FCs involve \emph{anterior cingulate and paracingulate gyri}, \emph{parahippocampal gyrus}, and \emph{hippocampus},
which complies with previous ASD-related findings~\cite{dichter2009autism,monk2009abnormalities,banker2021hippocampal}.
As shown in Fig.~\ref{fig_FCs}~(b), the identified discriminative brain ROIs in ANI identification include \emph{insula}, \emph{right temporal pole: superior temporal gyrus}, \emph{supplementary motor area}, and \emph{caudate nucleus}. 
These regions have also been reported in previous studies on HIV-related cognitive impairment~\cite{zhou2017motor,zhan2022resting,shin2017retrosplenial,chockanathan2019automated}.
These results further demonstrate the effectiveness of the BMR in detecting interpretable disease-associated biomarkers.
The identified discriminative brain ROIs
and FCs can be regarded as potential biomarkers to aid in clinical diagnosis.

\begin{figure}[t]
\setlength{\abovecaptionskip}{0pt}
\setlength{\belowcaptionskip}{-2pt}
\setlength{\abovedisplayskip}{-2pt}
\setlength{\belowdisplayskip}{-2pt}
\centering
\includegraphics[width=0.49\textwidth]{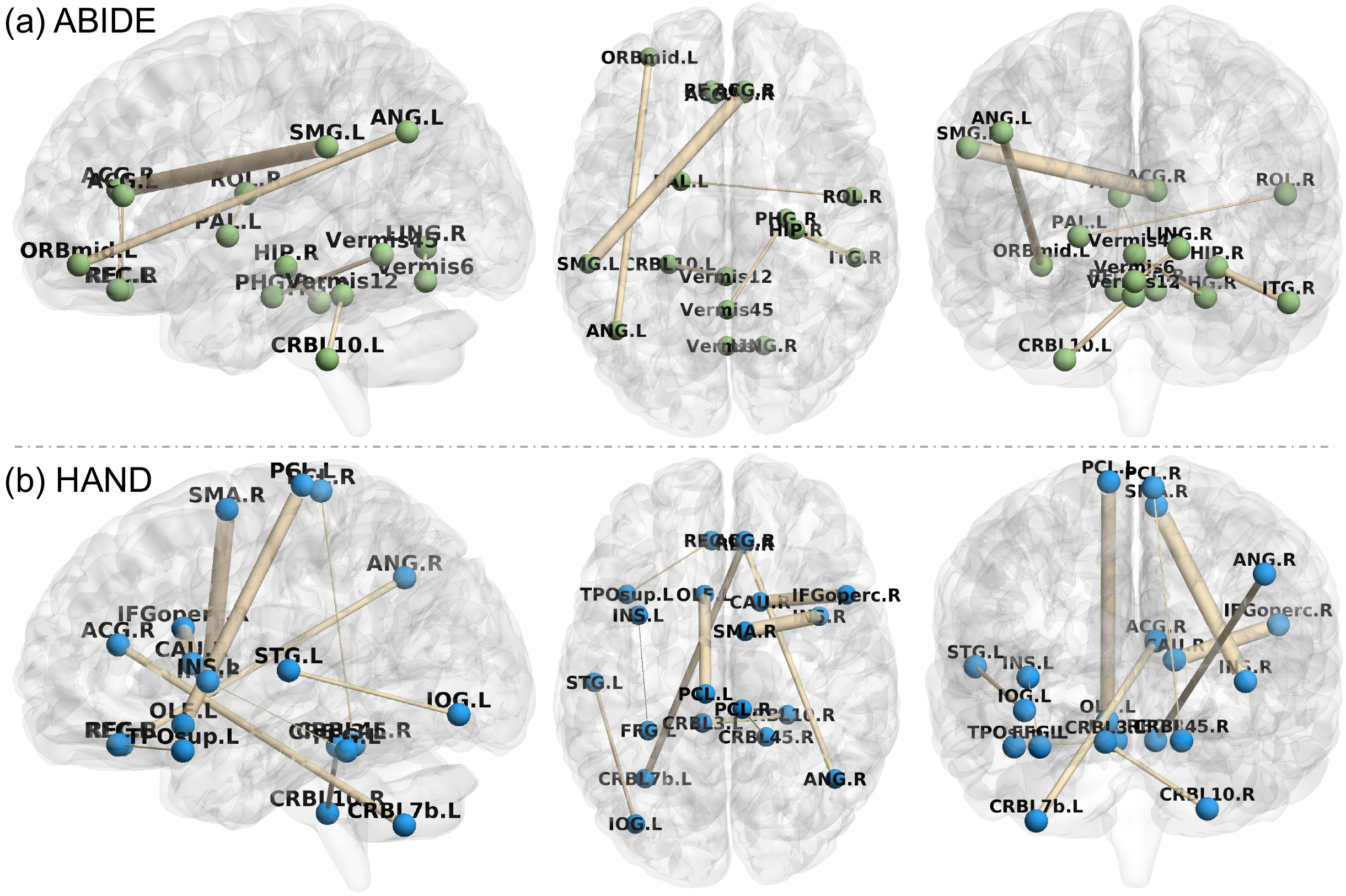}
\caption{Top ten discriminative functional connectivities identified by our BMR in (a) ASD vs. HC classification on NYU from ABIDE and (b) ANI vs. HC classification on HAND. }
\label{fig_FCs}
\end{figure}


\subsection{Influence of Hyperparameters}\label{parameter selection}
We have two hyperparameters (\ie, $\lambda_{1}$ and $\lambda_{2}$) in the proposed BMR (see Eq.~\eqref{loss}) to control contributions of brain modularity constraint and graph topology reconstruction constraint, respectively. 
To study their influences on the performance of BMR, we tune $\lambda_{1}$ and $\lambda_{2}$ within the range of $\{ 10^{-4}, 10^{-3}, \cdots, 10^1\}$ based on training and validation sets for ASD vs. HC classification on NYU from ABIDE, and report the results of BMR in Fig.~\ref{para}.  
It can be observed from Fig.~\ref{para} that the BMR with a large {{$\lambda_{1}$}} (\eg, $\lambda_{1}=10$) achieves worse performance.
The underlying reason may be that using a strong graph reconstruct constraint will make the model difficult to converge, thus degrading its learning performance. 
On the other hand, the BMR with a very weak modularity constraint (\eg, $\lambda_{2}=10^{-4}$) is generally inferior to that with relatively stronger modularity constraint (\eg, $\lambda_{2} =10^{-2}$).
These results mean that the BMR  can not achieve satisfactory performance when the BMR pays less attention to brain inherent modular structure, which further validates the effectiveness of the designed brain modularity constraint.
In particular, the BMR achieves the best AUC values with $\lambda_{1}=10^{-2}$  and $\lambda_{2}=10^{-2}$ in this task.


\begin{figure}[t]
\setlength{\abovecaptionskip}{0pt}
\setlength{\belowcaptionskip}{-2pt}
\setlength{\abovedisplayskip}{-2pt}
\setlength{\belowdisplayskip}{-2pt}
\centering
\includegraphics[width=0.45\textwidth]{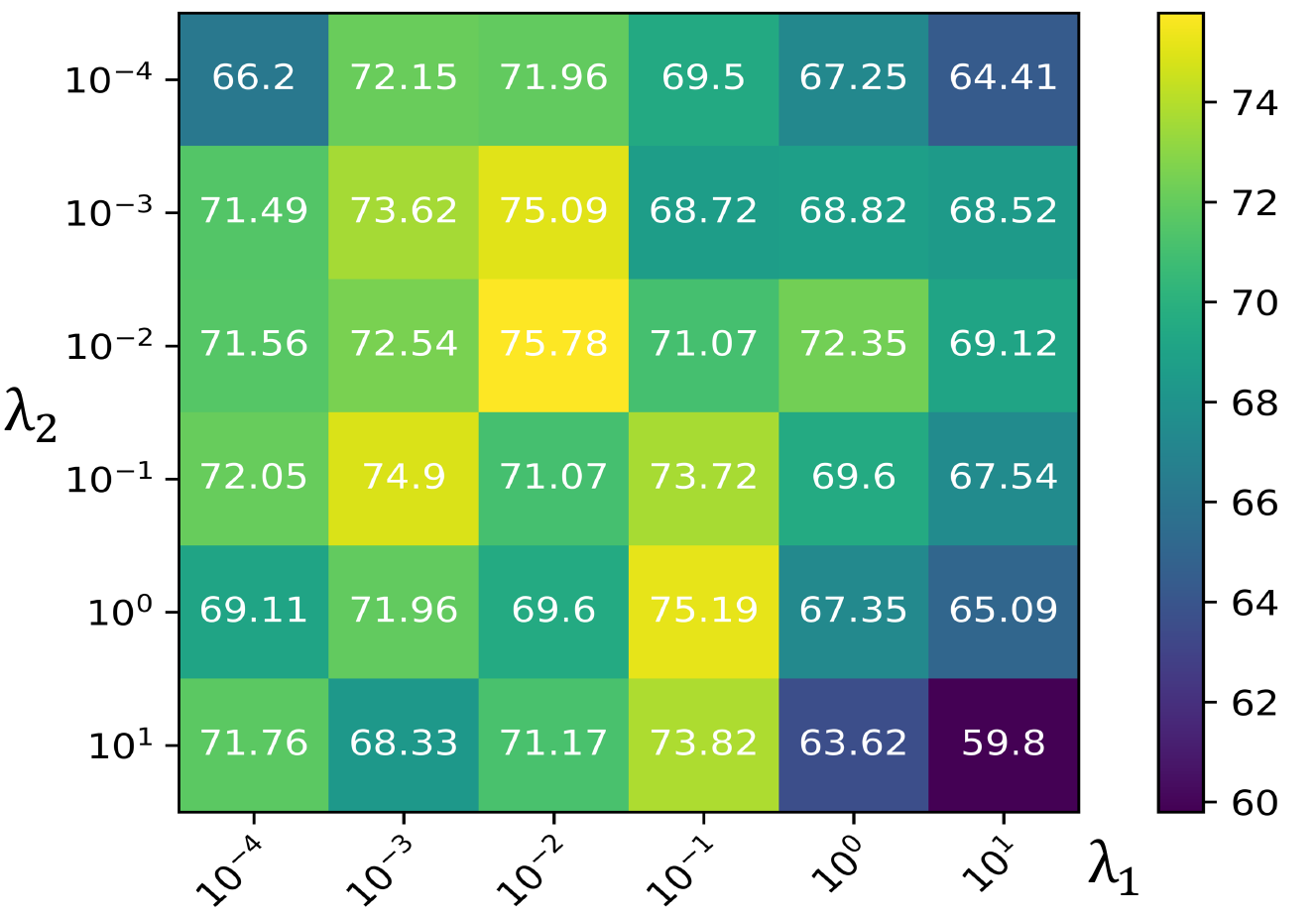}
\caption{AUC values (\%) of the proposed BMR under different hyperparameters (\ie, $\lambda_{1}$ and $\lambda_{2}$) in the task of ASD vs. HC classification on NYU site of ABIDE dataset.}
\label{para}
\end{figure}

\subsection{Influence of Spatial Feature Encoder}
In the main experiments, our BMR uses GAT as spatial feature encoder to capture dependencies among brain ROIs.
To investigate the influence of the spatial feature encoder, we replace GAT with the graph isomorphism network (GIN) to extract spatial fMI features in BMR, and call this variant as \textbf{BMR-GIN}. 
The results of BMR and BMR-GIN for ASD vs. HC classification on NYU are reported in Fig.~\ref{gnn}.  
It can be found from Fig.~\ref{gnn} that BMR achieves better performance than BMR-GIN in most cases. 
The main reason could be that,
compared with  BMR-GIN that treats neighboring nodes equally during the process of aggregating node features,
the BMR can adaptively assign different attention weights to different neighboring nodes so that the model can focus on important nodes,
thus improving learning performance.

\if false
The main reason could be that, compared with the BMR-GIN in which edge weights are pre-defined,
the BMR with GAT as spatial feature encoder can effectively model complex interactions between different  ROIs by adaptively learning edge weights with the help of multi-head attention mechanism, thus enhancing classification performance.
\fi

\begin{figure}[t]
\setlength{\abovecaptionskip}{0pt}
\setlength{\belowcaptionskip}{-2pt}
\setlength{\abovedisplayskip}{-2pt}
\setlength{\belowdisplayskip}{-2pt}
\centering
\includegraphics[width=0.49\textwidth]{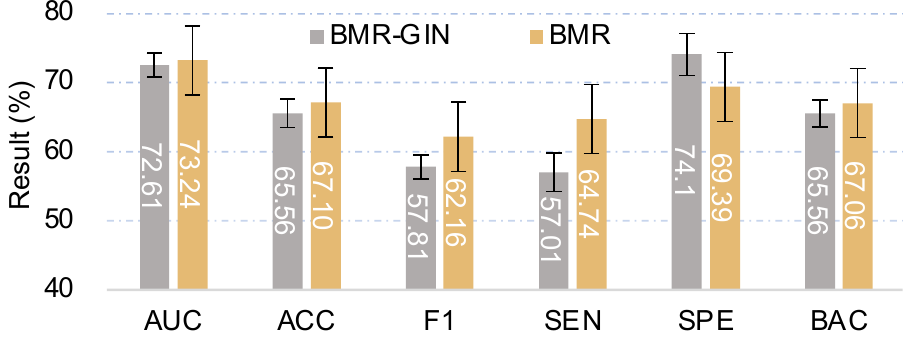}
\caption{Results of our BMR and its variant GMR-GIN (with GIN as spatial encoder) in ASD vs. HC classification on NYU.}
\label{gnn}
\end{figure}

\if false
{\color{cyan}
\subsection{Temporal Attention Localization}
For each subject, we can get a temporal attention matrix. 
Then, we can average each row of the matrix to obtain temporal point level weights.
So, we can compare temporal attention difference between two categories (\ie, patients and HCs)

Visualization via heat map
}
\fi
\begin{figure}[t]
\setlength{\abovecaptionskip}{0pt}
\setlength{\belowcaptionskip}{-2pt}
\setlength{\abovedisplayskip}{-2pt}
\setlength{\belowdisplayskip}{-2pt}
\centering
\includegraphics[width=0.49\textwidth]{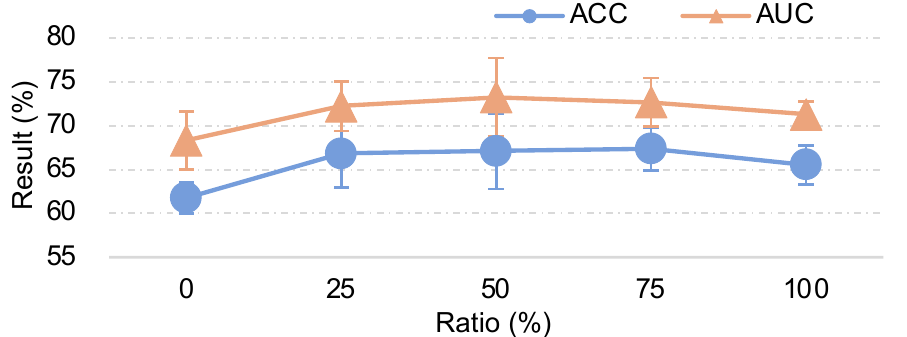}
\caption{Results of the proposed BMR with different modularity ratios in ASD vs. HC classification on NYU.}
\label{ratio}
\end{figure}
\subsection{Influence of Modularity Ratio}

In the proposed modularity constraint, we randomly select $m=50\%$ of all {{$\frac{N_{c}(N_{c}-1)}{2}$}} paired ROIs in the $c$-th module (with $N_{c}$ ROIs) to constrain the BMR.
To explore the influence of modularity ratio, 
we vary its values within the range of $\{0\%,25\%,\cdots,100\%\}$, and report the results on NYU site of ABIDE in Fig.~\ref{ratio}. 
As shown in Fig.~\ref{ratio}, with $m<75\%$, the ACC and AUC values of BMR generally improve as the increase of $m$. 
With a very large modularity ratio (\eg, $m=100\%$), BMR cannot achieve satisfactory performance. 
The possible reason is that using a too strong modularity constraint in BMR may lead to an over-smoothing problem, thus weakening the discriminative power of learned representations. 


\subsection{Influence of Sliding Window Size}

\begin{figure}[t]
\setlength{\abovecaptionskip}{0pt}
\setlength{\belowcaptionskip}{-2pt}
\setlength{\abovedisplayskip}{-2pt}
\setlength{\belowdisplayskip}{-2pt}
\centering
\includegraphics[width=0.48\textwidth]{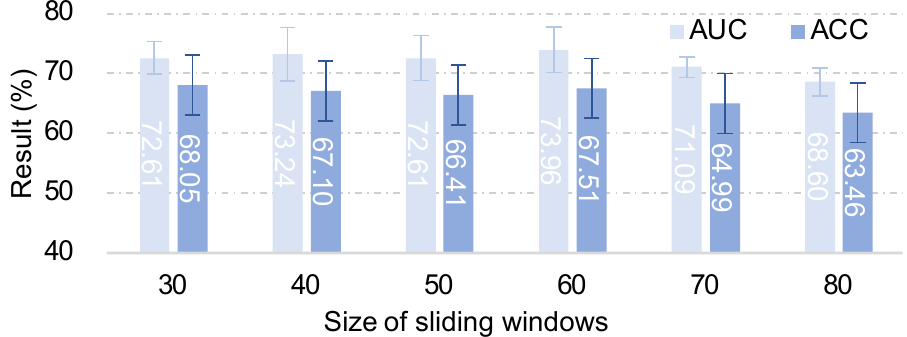}
\caption{Results of our BMR using different sizes of sliding windows in ASD vs. HC classification on NYU.}
\label{windowsize}
\end{figure}
In the main experiments, we use the sliding window strategy to generate dynamic BFNs with the window size of $\Gamma=40$.
To further explore the influence of sliding window size, 
we vary sliding window size within  $\{30,40,\cdots,80\}$ and record the results of BMR on NYU site of ABIDE in Fig.~\ref{windowsize}.  
As shown in Fig.~\ref{windowsize}, 
the BMR consistently yields promising performance (\ie, AUC $>72\%$) when the window size is within the range (\ie, $30\leq \Gamma \leq 60$).
But with large size of sliding windows (\eg, $\Gamma=80$), BMR cannot achieve good performance.
The reason could be that a larger window size provides lower temporal resolution, so that the BMR can not effectively capture temporal fluctuations within fMRI series.


\if false
\subsection{Influence of BFN Construction Algorithms}
{\color{red}
The input BFN of our BMR for each subject is constructed based on Pearson correlation (PC) coefficients between pairwise brain ROIs. 
To investigate the influence of different BFN construction methods, 
we also use sparse representation (SR)~\cite{} and low-rank representation (LR)~\cite{} for BFN construction, and denote these two variants as ``BMR-SR'' and ``BMR-LR'', respectively. 
The results of BMR and its two variants in ASD vs. HC 
 classification on NYU are reported in Table~\ref{FCN}.
From Table~\ref{FCN}, we can observe that  the BMR using PC 
outperforms its two variants (\ie, BMR-SR and BMR-LR) . 
%
The underlying reason could be that using PC to construct BFN can comprehensively model dependencies among regional BOLD signals, 
unlike the other two methods (\ie, BMR-SR and BMR-LR) that may overlook important functional connection information that are not explicitly represented.
}
 
\begin{table}[!tbp]
\setlength{\abovecaptionskip}{-0pt}
\setlength{\belowcaptionskip}{-2pt}
\setlength{\abovedisplayskip}{-2pt}
\setlength{\belowdisplayskip}{-2pt}
	\scriptsize
	\setlength{\tabcolsep}{1}
    \renewcommand{\arraystretch}{0.9}
	\caption{Performance of the BMR with different FC network construction strategies, including sparse representation (SR), low-rank
representation (LR) and Pearson correlation (PC), in ASD vs. HC 
 classification on NYU site of ABIDE. 
 Best results are shown in bold.  }
	\label{FCN}
	\centering
	\begin{tabular*}{0.48\textwidth}{@{\extracolsep{\fill}} l|cccccc}
		\toprule
        ~Method  & AUC (\%) & ACC (\%) & F1 (\%) & SEN (\%) & SPE (\%) & PRE (\%)   \\
        \midrule
              
        ~BMR-LR&64.89(2.55)&58.68(3.02)&51.75(3.46)&55.25(4.33)&63.12(1.92)&59.18(3.12)\\
        ~BMR-SR&67.14(5.27)&62.20(3.43)&50.41(11.63)&52.35(13.77)&\textbf{71.86(2.75)}&62.10(5.51)\\
       ~BMR&\textbf{73.24(4.48)}&\textbf{67.10(4.29)}&\textbf{62.16(4.53)}&\textbf{64.74(5.08)}&69.39(6.22)&\textbf{67.06(4.38)}\\
    
		\bottomrule
	\end{tabular*}
\end{table}
\fi

\if false
\begin{table*}[!t]
	\caption{Performance of the BMR with different FC network construction strategies, including sparse representation (SR), low-rank
representation (LR) and pearson correlation (PC). } 
	\label{network}
	\centering
	\setlength{\tabcolsep}{0.5pt}
    \begin{tabular*}
    {1\textwidth}{@{\extracolsep{\fill}}l|ccc| ccc| ccc| ccc}
    \toprule 
	\multirow{2}{*}{Method} & \multicolumn{3}{c|}{ASD vs. HC on NYU of ABIDE} &\multicolumn{3}{c|}{ASD vs. HC on UM of ABIDE}&\multicolumn{3}{c|}{ANI vs. HC on HAND}&\multicolumn{3}{c}{ICH vs. HC on HAND} \\
 
     \cmidrule(lr){2-4} \cmidrule(lr){5-7} \cmidrule(lr){8-10} \cmidrule(lr){11-13}
     & AUC (\%) & ACC (\%) & BAC (\%) & AUC (\%) & ACC (\%) & BAC (\%) & AUC (\%) & ACC (\%) & BAC (\%)& AUC (\%) & ACC (\%) & BAC (\%)
		 \\
		\midrule
         BMR-SR&62.2(3.5)&62.2(3.5)&62.2(3.5)&62.2(3.5)&62.2(3.5)&62.2(3.5)&62.2(3.5)&62.2(3.5)&62.2(3.5)&62.2(3.5)&62.2(3.5)&62.2(3.5)
         \\
        BMR-LR&
         \\
         BMR-PC&
         \\

		\bottomrule
	\end{tabular*}
\end{table*}
\fi

\subsection{Influence of Distance Metric}
We employ the cosine distance  in BMR to quantify the similarity between latent node-level representations within each module, as shown in Eq.~\ref{module constraint}. 
To study the effect of this metric, we compare BMR with its three variants: (1) \textbf{BMR-ED} with Euclidean distance, (2) \textbf{BMR-HD} with Hamming distance, and (3) \textbf{BMR-JD} with Jaccard similarity, with results reported in Table~\ref{distance}. 
\if false
Here, we further try to use different distance/similarity metrics (\ie, Euclidean distance, Hamming distance and
Jaccard similarity) to measure similarity between node representations and report results in Table~\ref{distance}.
\fi 
As shown in Table~\ref{distance}, BMR using four different distance metrics in the modularity constraint achieves comparable results. 
This implies that our BMR is not sensitive to the distance metrics used in the modularity constraint. 
\if false
the BMR achieves slightly better performance than three variants (\ie, BMR-ED, BMR-HD and BMR-JS).
Overall, the BMR using different distance metrics in the modularity constraint achieves comparable results, which means the BMR is insensitive to distance metrics.
\fi 
 
\if false 
\begin{table*}[!t]
	\caption{Performance of the BMR with different distance metrics in modularity constraint, including Euclidean distance (ED), Hamming distance (HD),
Jaccard similarity (JS) and Cosine similarity (CS). } 
	\label{distance}
	\centering
	\setlength{\tabcolsep}{0.5pt}
    \begin{tabular*}
    {1\textwidth}{@{\extracolsep{\fill}}l|ccc| ccc| ccc| ccc}
    \toprule 
	\multirow{2}{*}{Method} & \multicolumn{3}{c|}{ASD vs. HC on NYU of ABIDE} &\multicolumn{3}{c|}{ASD vs. HC on UM of ABIDE}&\multicolumn{3}{c|}{ANI vs. HC on HAND}&\multicolumn{3}{c}{ICH vs. HC on HAND} \\
 
     \cmidrule(lr){2-4} \cmidrule(lr){5-7} \cmidrule(lr){8-10} \cmidrule(lr){11-13}
     & AUC (\%) & ACC (\%) & BAC (\%) & AUC (\%) & ACC (\%) & BAC (\%) & AUC (\%) & ACC (\%) & BAC (\%)& AUC (\%) & ACC (\%) & BAC (\%)
		 \\
		\midrule
         BMR-ED&62.2(3.5)&62.2(3.5)&62.2(3.5)&62.2(3.5)&62.2(3.5)&62.2(3.5)&62.2(3.5)&62.2(3.5)&62.2(3.5)&62.2(3.5)&62.2(3.5)&62.2(3.5)
         \\
        BMR-HD&
         \\
         BMR-JS&
         \\
         BMR-CS&
         \\

		\bottomrule
	\end{tabular*}
\end{table*}
\fi 

\begin{table}[!tbp]
\setlength{\abovecaptionskip}{-0pt}
\setlength{\belowcaptionskip}{-0pt}
\setlength{\abovedisplayskip}{-2pt}
\setlength{\belowdisplayskip}{-2pt}
\scriptsize
\setlength{\tabcolsep}{1pt}
\renewcommand{\arraystretch}{0.9}
\caption{Performance of the proposed BMR and its three variants that use different similarity metrics in the proposed modularity constraint in ASD vs. HC classification on NYU.}
\label{distance}
\centering
\begin{tabular*}{0.48\textwidth}{@{\extracolsep{\fill}} l|cccccc}
\toprule
Method  & AUC (\%) & ACC (\%) & F1 (\%) & SEN (\%) & SPE (\%) & PRE (\%)   \\
\midrule
BMR-ED & 73.00(1.69) & 65.88(2.86) & 60.62(3.33) & 64.37(3.80) & 68.19(2.51) & 66.28(2.40) \\
BMR-HD & 72.03(2.69) & 64.75(2.57) & 59.04(3.21) & 60.88(4.21) & 68.25(4.20) & 64.57(2.52) \\
BMR-JS & 73.12(3.01) & 66.92(4.54) & 61.48(4.85) & 63.13(4.80) & \textbf{70.53(6.20)} & 66.83(4.61) \\
BMR    & \textbf{73.24(4.48)} & \textbf{67.10(4.29)} & \textbf{62.16(4.53)} & \textbf{64.74(5.08)} & 69.39(6.22) & \textbf{67.06(4.38)} \\
\bottomrule
\end{tabular*}
\end{table}

              



\subsection{Influence of BFN Sparsity Ratio}
Following~\cite{kim2021learning}, we empirically retain the top 30\% strongest edges (\ie, sparsity ratio) in each BFN in the experiments.
To study the impact of sparsity ratio,  
we vary its value within $\{10\%,\cdots,100\%\}$ and report AUC and ACC values in ASD vs. HC classification on NYU in Fig.~\ref{sparsity}.
As shown in Fig.~\ref{sparsity},
our BMR achieves stable results when sparsity ratio is $<60\%$.
For example, BMR obtains the AUC values of $73.69\%$ and $73.32\%$ when sparsity ratios are set as $10\%$ and $50\%$, respectively. 
The reason could be that BMR retains the most reliable and informative connections in BFNs by prioritizing the strongest edges, 
reducing the impact of noisy or redundant connections. 
But when the sparsity rate is large (\eg, $>90\%$), the BMR cannot produce good results. 
The possible reason is that the BFN with such large sparsity can not effectively reflect topology information 
due to the loss of too many connections.

\begin{figure}[t]
\setlength{\abovecaptionskip}{0pt}
\setlength{\belowcaptionskip}{-2pt}
\setlength{\abovedisplayskip}{-2pt}
\setlength{\belowdisplayskip}{-2pt}
\centering
\includegraphics[width=0.48\textwidth]{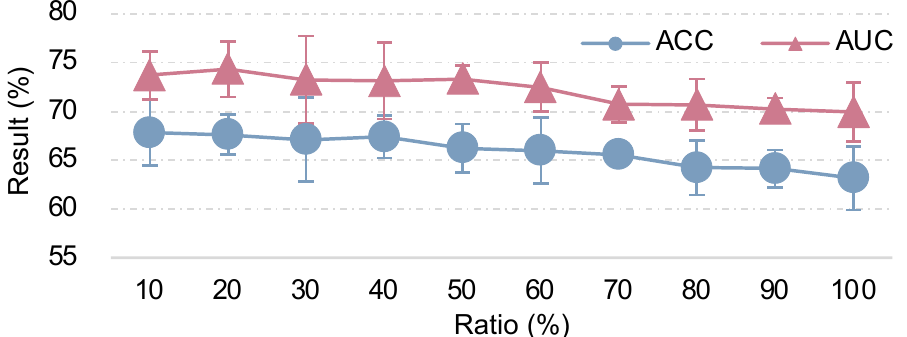}
\caption{Results of the proposed BMR with different sparsity ratios of BFN in ASD vs. HC classification on NYU.}
\label{sparsity}
\end{figure}
\subsection{Limitations and Future Work}
Several limitations need to be considered in future work. 
\emph{First},
we only characterize pairwise relationships of ROIs within three prominent neurocognitive modules (\ie,  SN, CEN, and DMN) as prior knowledge to design the modularity constraint in BMR.
It is meaningful to design disease-specific modularity constraints based on neurocognitive research and clinical experience in the future.
\emph{Second},
the BMR incorporates known brain modular organization into fMRI representation learning.
Future work will seek to design new algorithms that can automatically detect unknown brain modular structures during graph/BFN learning to characterize disease-induced brain changes.
\emph{Besides}, 
the BMR needs to be trained on labeled fMRI data in a supervised manner. 
In future work, 
we will employ unsupervised contrastive learning strategies~\cite{wang2022contrastive} to pre-train the feature encoder on large-scale unlabeled data to learn more discriminative fMRI features. 

\if false
{{\color{blue}1. Using disease-specific modularity; 2. data section & inter-site heterogeneity; 3. }
\fi

\section{Conclusion}
In this paper, we propose a Brain Modularity-constrained dynamic Representation learning (BMR) framework for interpretable fMRI analysis.
Specifically, we first construct a dynamic graph/BFN for each subject, and then design a brain modularity-constrained GNN model for dynamic graph representation learning, where a novel modularity constraint is developed to encourage nodes within the same module to share similar embeddings. 
\if false
By exploiting brain modularity prior, the learned dynamic representation can capture intrinsic modular organization in the brain, thus enhancing the interpretability of the BMR.
\fi 
We also propose a graph topology reconstruction constraint to preserve original topology information of input BFNs during representation learning.
Finally, we perform brain disorder prediction and biomarker detection by analyzing disease-related functional connectivities and brain regions, aiming to provide biological evidence for clinical practice. 
Extensive experiments demonstrate the effectiveness of BMR in fMRI-based brain disorder detection.

\vspace{-4pt}
\bibliographystyle{model2-names.bst}
\biboptions{authoryear}
\bibliography{refs}

\end{document}